%% file: main.tex
\documentclass[10pt,conference]{IEEEtran}

\IEEEoverridecommandlockouts
\def\BibTeX{{\rm B\kern-.05em{\sc i\kern-.025em b}\kern-.08em
    T\kern-.1667em\lower.7ex\hbox{E}\kern-.125emX}}
\usepackage{multirow}
\usepackage{url}
\usepackage{tcolorbox}
\usepackage{booktabs}
\usepackage{cite}
\usepackage{amsmath,amssymb,amsfonts}
\usepackage{algorithmic}
\usepackage{graphicx}
\usepackage{textcomp}
\usepackage{xcolor}

\newcommand{\tool}{DualSC}

\begin{document}

\title{{\tool}: Automatic Generation and Summarization of Shellcode via Transformer and Dual Learning}

\author{\IEEEauthorblockN{Guang Yang\IEEEauthorrefmark{2},  
Xiang Chen\IEEEauthorrefmark{2}\IEEEauthorrefmark{3}\IEEEauthorrefmark{1},  
Yanlin Zhou\IEEEauthorrefmark{2}, 
Chi Yu\IEEEauthorrefmark{2}}
\IEEEauthorblockA{\IEEEauthorrefmark{2}\textit{School of Information Science and Technology},
\textit{Nantong University}, China\\
\IEEEauthorrefmark{3}\textit{State Key Laboratory of Information Security},\textit{ Institute of Information Engineering, Chinese Academy of Sciences}, China\\
Email: novelyg@outlook.com, xchencs@ntu.edu.cn, 1159615215@qq.com, yc\_struggle@163.com}
}

% \author{Anonymous Author(s)}

\maketitle

\begingroup
\renewcommand{\thefootnote}{}
\footnotetext[1]{\IEEEauthorrefmark{1} Xiang Chen is the corresponding author.}
\endgroup

\begin{abstract}
A shellcode is a small piece of code and it is executed to exploit a software vulnerability, which allows the target computer to execute arbitrary commands from the attacker through a code injection attack.
Similar to the purpose of automated vulnerability generation techniques, the automated generation of shellcode can generate attack instructions, which can be used to detect vulnerabilities and implement defensive measures. While the automated summarization of shellcode can help users unfamiliar with shellcode and network information security understand the intent of shellcode attacks.
In this study, we propose a novel approach {\tool} to solve the automatic shellcode generation and summarization tasks. Specifically, we formalize automatic shellcode generation and summarization as dual tasks, use a shallow Transformer for model construction, and design a normalization method Adjust\_QKNorm to adapt these low-resource tasks (i.e., insufficient training data). Finally, to alleviate the out-of-vocabulary problem, we propose a rule-based repair component to improve the performance of  automatic shellcode generation. In our empirical study,
we select a high-quality corpus Shellcode\_IA32 as our empirical subject. This corpus was gathered from two real-world  projects based on the line-by-line granularity.
We first compare {\tool} with six state-of-the-art baselines from the code generation and code summarization domains in terms of four performance measures. The comparison results show the competitiveness of {\tool}. Then, we verify the effectiveness of the component setting in {\tool}.
Finally, we conduct a human study to further verify the effectiveness of {\tool}.

\end{abstract}


\input{1introduction}
\input{2background}

\input{3approach}
\input{4setup}

\input{5result}

\input{6discuss}
\input{7threats}
\input{8conclusion}

\section*{Acknowledgment}

The authors would like to thank the anonymous reviewers for their insightful comments and suggestions. 
This work is supported in part by the National Natural Science Foundation of
China (Grant no. 61872263), The Open Project of State Key Laboratory of Information Security (Institute of Information Engineering, Chinese Academy of Sciences) (Grant No. 2020-MS-07).
\bibliographystyle{IEEEtran}
\bibliography{mylib}
\end{document}

%% file: 1introduction.tex
\section{Introduction}
\label{sec:intro}

A shellcode is a small piece of code that is executed to exploit a software vulnerability. Specifically, shellcode means writing code that returns a remote shell when executed. The shellcode is usually written via assembly instructions, which makes the target computer execute arbitrary instructions from the attacker through a code injection attack, so many hackers in the hacking community refer to the payload portion of a code injection attack as the shellcode~\cite{mason2009english}. 
Shellcode is considered a key element of a security attack, and malicious shellcodes can perform DDoS attacks, data theft, and run malware against the target systems~\cite{arce2004ShellCode, wang2010attack, park2018classification}. 
However, shellcode with good intentions can be used to identify vulnerabilities and then repair systems by fixing these vulnerabilities~\cite{polychronakis2010comprehensive, ding2017accurate, bao2017your, shangru2019survey}. 
% \cx{Shellcode takes the form of directly executable machine code and then presents many defense measures, which attempt to detect its presence or prevent its execution. }
Therefore, researchers expected to use shellcode as the tool, which can be exploited to find security vulnerabilities in the software, and automated shellcode generation techniques have become an  active research topic~\cite{arce2004ShellCode,borders2007spector,mason2009english,chen2011automatic,bao2017your,patel2020automatic,liguori2021shellcode_ia32}.

\begin{figure}[htbp]
	\centering
    \vspace{-1mm}
	\includegraphics[width=0.5\textwidth]{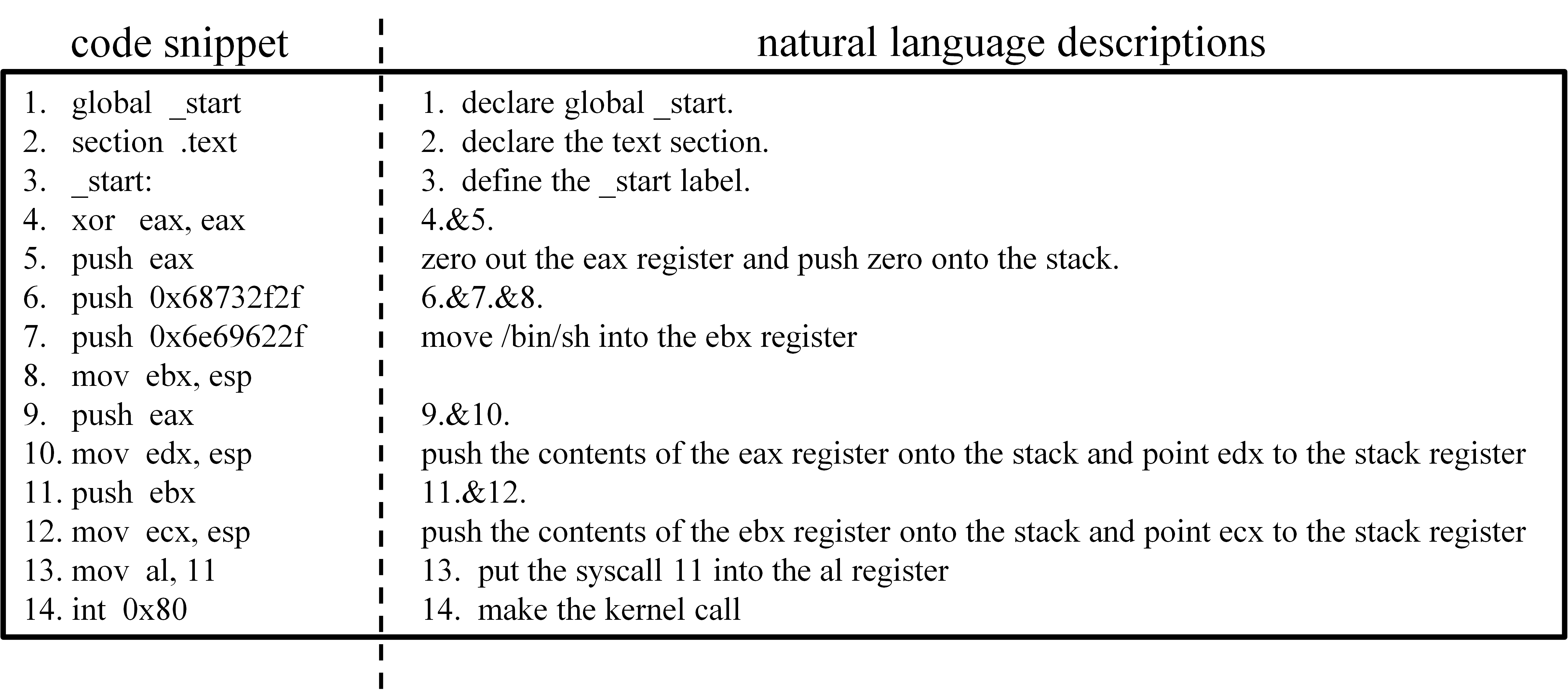}
	\caption{The example of the code snippet and its corresponding comments written in English from the corpus shellcode\_IA32}
    \vspace{-1mm}
	\label{fig:snippet}
\end{figure}

However, automatic shellcode generation is a challenging task as they are usually written in assembly language and the most complex shellcode can run into the assembly code with hundreds of lines. Moreover, it is also challenging to summarize the functional description of shellcodes, since identifying the attacking intent of these shellcodes can be difficult for users who are not familiar with shellcode and network information security. \figurename~\ref{fig:snippet} shows an example of a code snippet in our used corpus shellcode\_IA32. In this figure, the code snippet is on the left and the summarization (i.e., the code comment) based on the natural language description is on the right. Notice in this code snippet,  multiple lines of code may correspond to one comment since some instructions do not make sense when considered in isolation, which can make the shellcode summarization more challenging. To construct a parallel corpus (i.e., a one-to-one mapping between code and its corresponding comment), shellcode\_IA32 break lines by ``\textbackslash{}n". Based on the characteristic of this corpus, we are concerned with the shellcode generation and summarization based on the line-by-line granularity in this study.
Moreover, due to the small size of this corpus, automatic shellcode generation and summarization can be treated as low-resource tasks, which makes solving these two tasks more challenging.

Automated approaches to code generation and summarization tasks are of great importance for software development and maintenance~\cite{de2005study}. The generated high-quality comments and code can help to improve developer productivity and then improve software quality~\cite{xia2017measuring}. However, generating high-quality code or comments is time-consuming and error-prone. Previous studies~\cite{feng2020codebert,phan2021cotext,wang2021codet5,ahmad2021unified,chirkova2021empirical} on automated code generation and 
code summarization mainly focused on code snippets for specific programming languages (such as Java, Python, C\#). Recently, there are also some studies, which investigate code generation and  summarization  for specific software artifacts (such as Bash, SQL, and smart contract)~\cite{lin2018nl2bash, zavershynskyi2018naps,yang2021multi,yang2021ccgir}. To the best of our knowledge, there is only a study~\cite{liguori2021shellcode_ia32} on automatic shellcode generation and no studies focus on automatic shellcode summarization.

In this study, we are the first to formalize the automatic shellcode generation and summarization as a dual learning problem. Then we propose a fully data-driven approach {\tool} to solve the above two dual tasks. Moreover, we design a simple yet effective rule-based repair component to improve the performance of the code generation task. 
Since the automatic shellcode generation and summarization tasks are performed on insufficient training data (i.e., the low-resource tasks) and these two tasks have a strong correlation, we exploit the symmetric structure (i.e., code $\Leftrightarrow$ comment) 
% and linguistic properties (i.e., both are written in English and have a certain naturalness~\cite{allamanis2018survey}) 
between the two tasks via dual learning and use a shallow Transformer model to learn them simultaneously. This setting can increase the available training data for model construction. Moreover, it can also improve the generalization ability by learning the commonalities between these two tasks and transferring knowledge between these two tasks through shared representation. 
Finally, the method QKNorm~\cite{henry2020query} converts the  self-attention computation from dot product to cosine similarity as a way to adapt Transformer to low resource tasks. However, cosine similarity mainly focuses on direction-based similarity (i.e., the closer the direction between two vectors, the higher the value of cosine similarity). Therefore QKNorm is not sensitive to numerical differences between two vectors. To take the numerical difference information into account in the self-attention computation, we design the method Adjust\_QKNorm.

To verify the effectiveness of our proposed approach {\tool}, we chose the high-quality corpus shellcode\_IA32~\cite{liguori2021shellcode_ia32} as our experimental subject. We compared {\tool} with six state-of-the-art baselines. Specifically, the first three baselines are selected from the information retrieval domain and the remaining three baselines are selected from the neural network domain (including the recent approach for the shellcode generation task~\cite{liguori2021shellcode_ia32}). Then we evaluated the performance of code generation in terms of three performance measures (i.e., BLEU, Rouge-L, and Accuracy) and the performance of code summarization in terms of three performance measures (i.e., BLEU, Rouge-L, and METEOR). The empirical results show that {\tool} can outperform these baselines. Finally, we conduct a human study to further verify the effectiveness of our proposed approach.

In summary, the main contributions of our study can be summarized as follows.

\begin{itemize}
  \item We formalize the automatic generation and summarization of shellcode as the dual tasks. Then we learn these two tasks simultaneously via the shallow Transformer and dual learning, which can share knowledge and then improve the performance and generalization of the trained models.
  
  \item We choose the corpus from real-world projects as our experimental subject. Final comparison results with six state-of-the-art baselines show the competitiveness of {\tool} in terms of the corresponding performance measures both in the automatic generation and summarization tasks of shellcode. 

  \item We share our scripts, the trained models, and the used corpus\footnote{\url{https://github.com/NTDXYG/DualSC}} to facilitate the replication of our study and encourage more follow-up studies on this research topic.
\end{itemize}

%% file: 2background.tex
\section{Related Work}
\label{sec:related}

In this section, we first summarize the related work for shellcode analysis.
Then we summarize the related work for code generation and code summarization.  Finally, we emphasize the novelty of our study.

\subsection{Shellcode Analysis}

Arce~\cite{arce2004ShellCode} was the first to describe the development of shellcode and introduced the working principle and defensive measures of shellcode. 
Borders et al.~\cite{borders2007spector} proposed Spector. This approach used symbolic execution to extract meaningful API calls and their arguments from shellcode, which can help the users know what the vulnerability is used for. 
% Spector could help classify different payloads that have the same behavior. 
Mason et al.~\cite{mason2009english} proposed a technique for automatically generating shellcode using natural language generation techniques.
% , which can convert arbitrary shellcode into a representation \cx{that superficially resembles English prose}.  
Chen et al.~\cite{chen2011automatic} designed a new type of shellcode, which is based on JOP (Jump-Oriented Programming) gadgets. Experimental results proved that this new JOP-based shellcode can bypass most existing ROP (Return-Oriented Programming) defenses. 
Bao et al.~\cite{bao2017your} designed  an automated shellcode porting system ShellSwap for remote exploits. This system is the first end-to-end system, which can modify the observed vulnerabilities and replace the original shellcode with an arbitrary replacement shellcode. 
Patel et al.~\cite{patel2020automatic} treated the writing of printable shellcodes as an important task. Then they proposed a new encoding algorithm that used looped decoding to reduce the size of these automatically generated shellcodes. 
Liguori et al.~\cite{liguori2021shellcode_ia32} constructed a corpus Shellcode\_IA32 from two real-world projects and made the first attempt to use neural machine translation (NMT) to solve the task of automated shellcode generation. This corpus contains a total of 3,200 pairs of instructions from assembly language code snippets and their corresponding comments described in English. To our best knowledge, this is the largest corpus for shellcode at present.

\subsection{Code Generation and Code Summarization}

Code generation task is an active topic in the field of mining software repository. Recently,  researchers mainly applied deep learning to code generation.
Rabinovich et al.~\cite{rabinovich2017abstract} introduced abstract syntax networks, which included modular encoder-decoder architecture that can readily accommodate a variety of tasks with structured output spaces. 
Yin et al.~\cite{yin2017syntactic} proposed a novel neural architecture powered by a grammar model to explicitly capture the target syntax as prior knowledge. Yin and Neubig~\cite{yin2018tranx} presented TRANX, a transition-based neural semantic parser that maps natural language (NL) utterances into formal meaning representations (MRs). 
Hayati et al.~\cite{hayati2018retrieval} introduced a method RECODE, which makes
it possible to explicitly reference existing code examples within a neural code generation model.
Agashe et al.~\cite{agashe2019juice} presented JuICe, a corpus of 1.5 million examples with a curated test set of 3.7K instances based on online programming assignments, to study the code generation task. 
Sun et al.~\cite{sun2019grammar} proposed a grammar-based structural convolutional neural network for code generation, which generates a program by predicting the grammar rules of the programming language. 
Xu et al.~\cite{xu2020incorporating} explored the effectiveness of incorporating two varieties of external knowledge into NL-to-code generation: automatically mined NL-code pairs from the online programming QA forum Stack Overflow and programming language API documentation. 
Zhong et al.~\cite{zhong2020semantic} proposed a method for program generation
based on semantic scaffolds, which are lightweight structures and can represent the high-level semantic and syntactic composition of a program. 
Sun and Zhu~\cite{sun2020treegen} proposed  a novel tree-based neural architecture TreeGen, which uses the attention mechanism of Transformers to alleviate the long dependency problem, and introduces a novel AST reader (i.e., encoder) to incorporate grammar rules and AST structures into the network.

Recently, most of the previous studies on code summarization followed deep learning-based approaches (i.e., encoder-decoder framework). 
For example, Iyer et al.~\cite{iyer2016summarizing} first proposed an approach code-NN via an attention-based neural network.
Allamanis et al.~\cite{allamanis2016convolutional} proposed a model in which the encoder uses CNN and the attention mechanism, and the decoder uses GRU.
Liang and Zhu~\cite{liang2018automatic} used Code-RNN to encode the source code into the vectors, and then they used Code-GRU to decode the vectors to code comments.
Hu et al.~\cite{hu2018deep} proposed the approach DeepCom by analyzing abstract syntax trees (ASTs). To better present the structure of ASTs, they proposed a new structure-based traversal (SBT) method.
Later, Hu et al.~\cite{hu2020deep} further proposed the improved approach Hybrid-DeepCom.  For example, the identifiers satisfying the Camel casing naming convention are split into multiple words to alleviate the OOV problem.
Yang et al.~\cite{yang2021comformer} proposed a novel method ComFormer based on Transformer and fusion method-based hybrid code presentation.
Recently, Kang et al.~\cite{kang2019assessing} investigated whether using the pre-trained word embedding could improve the model performance. They surprisingly found that using the pre-trained word embedding based on code2vec~\cite{alon2019code2vec} or Glove~\cite{pennington2014glove} could not help to improve the performance.
Leclair et al.~\cite{leclair2019neural} proposed the method ast-attendgru, which combines words from code and code structure.
Chen et al.~\cite{chen2018neural} proposed a neural framework, which allows bidirectional mapping between a code retrieval task and a code comment generation task. 
Wei et al.~\cite{wei2019code} also utilized the correlation between the source code summarization task and the code generation task. Then they proposed a dual training framework.
Ye et al.~\cite{ye2020leveraging} exploited the probabilistic correlation between the source code summarization task and the code generation task via dual learning.
Ahmad et al.~\cite{ahmad2020transformer} used the Transformer model to generate code comments. In their study, they combined the self-attention and the copy attention as the attention mechanism of the model and analyzed the influence of the absolute position and pairwise relationship on the performance of source code summarization.
Liu et al.~\cite{liu2020atom} proposed the ATOM method. This method uses a hybrid sorting module, which can select the most relevant commit messages from those that have been generated or retrieved   iteratively.

\subsection{Novelty of Our Study}

In our study, we propose a fully data-driven approach {\tool} to solve the shellcode generation and shellcode summarization tasks. Considering the characteristics of automatic shellcode generation and summarization tasks, we formalize them as dual tasks and then use a shallow Transformer for simultaneous learning. Moreover, we also propose a normalization method Adjust\_QKNorm for both tasks to accommodate {\tool}  in the low-resource scenario. Finally, to alleviate the OOV problem, we propose a rule-based repair component to improve the accuracy of {\tool} on automatic shellcode generation.

%% file: 3approach.tex
\section{Our Proposed Approach {\tool}}
\label{sec:method}

In this section, we first show the framework of {\tool}.
Then we show the details of model architecture in {\tool}.
Finally, we introduce the details of our designed rule-based repair component in {\tool}.

\subsection{Framework of {\tool}}

\begin{figure*}[htbp]
	\centering
  \vspace{-1mm}
	\includegraphics[width=0.85\textwidth]{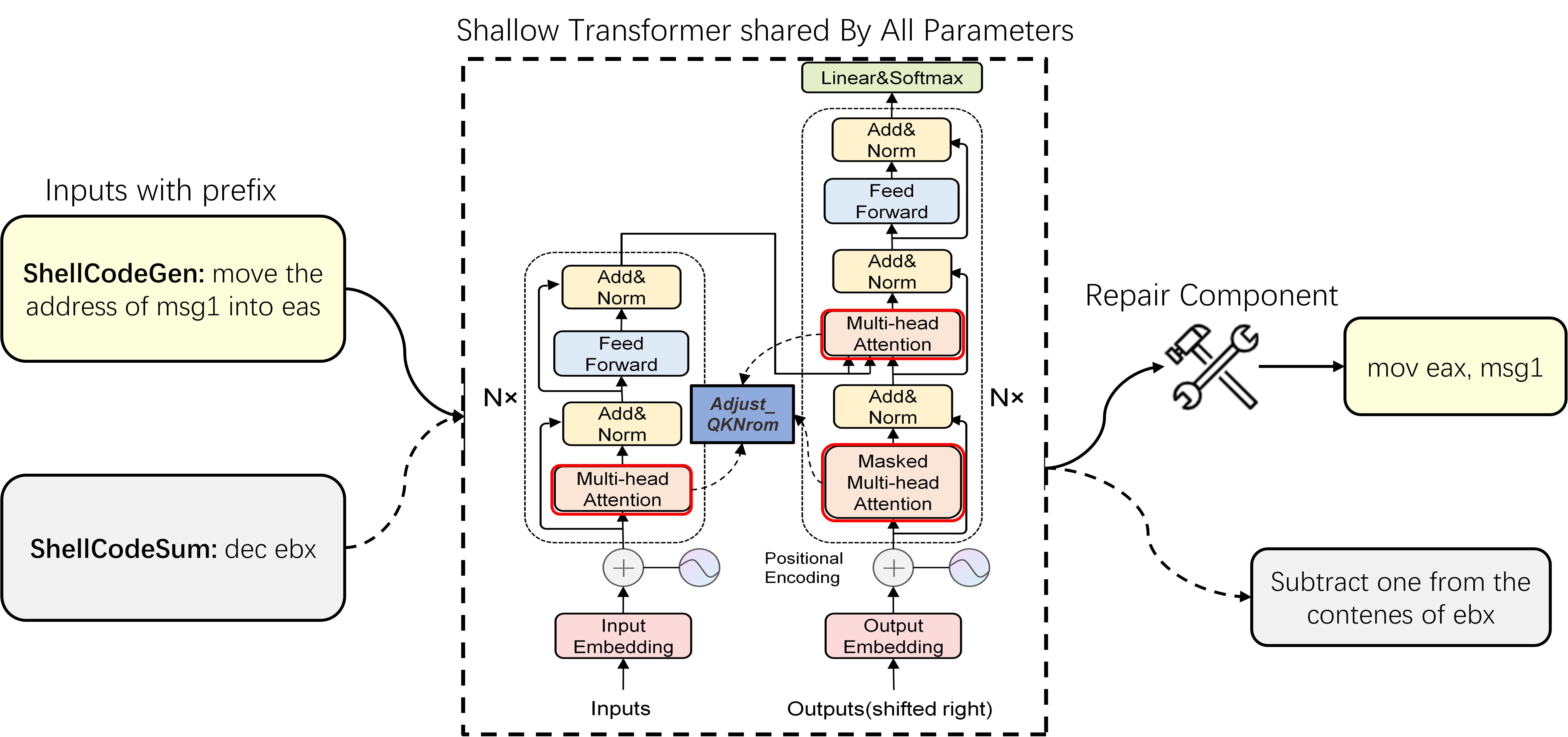}
	\caption{Framework of our proposed approach {\tool}}
  \vspace{-1mm}
	\label{fig:framework}
\end{figure*}

The overall framework of {\tool} is shown in \figurename~\ref{fig:framework}. 
% Unlike traditional dual learning~\cite{he2016dual}, shellcode's code and comments are written in English and there is the naturalness in both code and comments. 
Inspired by the model T5~\cite{raffel2019exploring}, we use the same model to learn both the primal task and the dual task. However, due to the limitation of GPU resources, 
% we only consider solving the automatic shellcode generation task and the automatic shellcode summarization task with limited GPU resources. Therefore 
we use a shallow Transformer model~\cite{van2020optimal}, which can learn with limited GPU resources. We also improve the Self-Attention computation~\cite{vaswani2017attention} by proposing Adjust\_QKNorm to adapt the shallow Transformer in the low-resource scenario. The key idea of {\tool} is to exploit the duality between the primal task (i.e., mapping from the domain $X$ to the domain $Y$) and the dual task (i.e., mapping from the domain $Y$ to the domain $X$) to improve the performance of both tasks. Before the data is input into the model, it is distinguished by a Prompt paradigm~\cite{liu2021pre}, where each specific task is distinguished by a prefix. Specifically, for the automatic shellcode generation task, we add the prefix ``ShellCodeGen:" for its input data. For the automatic shellcode summarization task, we add the prefix  ``ShellCodeSum:" for its input data. During the training and decoding, we share all parameters of the model to improve its performance and generalization ability. Finally, we manually analyzed the examples with the error output for the automatic shellcode generation task and designed a rule-based repair component to further improve the quality of the generated shellcode.

\subsection{Shallow Transformer-based Model}

Through the Embedding layer and Positional Encoding, Transformer can covert the input sequence $X_{input}$ = $( prefix, x_{1}, x_{2}, \cdots, x_{n} )$ to the vector $X_{emb}$ and feed this vector to the Encoder Layer. In the Encoder part, Transformer inputs the vector $X_{emb}$ into the multi-head attention layer. The attention function is computed simultaneously based on three matrices: $Q$ (queries), $K$ (keys), and $V$ (values).

\begin{equation}
Q, K, V=Linear(X_{emb})
\end{equation}

\begin{equation}
X_{atten}=Attention(Q, K, V)
\label{self-attention}
\end{equation}

In the classical Transformer, self-attention is calculated in relation to queries ($Q$), keys ($K$), and values ($V$). The dot product between queries and keys is first calculated, then each is given a division by $\sqrt{d_{k}}$ and the $\operatorname{softmax}$ function is applied to obtain the weights of the corresponding values.
\begin{equation}
\text { Attention }(Q, K, V)=\operatorname{softmax}\left(\frac{Q K^{T}}{\sqrt{d_{k}}}\right) V.
\end{equation}

Vaswani et al.~\cite{vaswani2017attention} investigated the two computational approaches (i.e., additive attention and dot-product), and analyzed the differences between them. With the advances of the Transformer,  researchers conducted various architectural and functional modifications to improve its performance on low-resource tasks, including reducing the depth of the model, adding a regularization penalty or adjusting the order of the LayerNorm~\cite{van2020optimal,xiong2020layer,xu2019understanding,nguyen2019transformers, shen2020powernorm}. The method QKNorm proposed by Henry et al.~\cite{henry2020query} can achieve promising results on low-resource machine translation tasks.  Specifically, the method QKNorm first performed L2 normalization of $Q$ and $K$ before performing the calculation, at which point the result obtained from the dot product is expressed as a cosine similarity calculation of $Q$ and $K$. Thus the calculation results can be controlled  in the interval [-1, 1], and does not need to be divided by $\sqrt{d_{k}}$. Instead, the model can be trained by multiplying with a learnable parameter $g$ for training. 

% QKNorm的改进被证实了在低资源任务下的有效性，它的主要动机是将矩阵Q和K的相似度计算从向量点积改进为向量间的余弦相似度。 这就导致了一个问题，余弦相似度主要侧重于两个向量之间的“方向相似性”，即两个向量之间的方向越相近，余弦相似度的值就越高。它对于向量之间数值的大小差异并不敏感。为了将向量的数值大小差异信息也考虑到self-attention计算中，我们提出了Adjust\_QKNorm。考虑到Q和K矩阵是三维向量，第一个维度意为batch_size的大小，第二个维度意为输入序列的长度，第三个维度意为单词对应的嵌入的表征向量。因此，我们对于Q和K矩阵的最后一个维度的值都减去它们对应的平均值，这会导致最后一个维度上所有元素的和等于零，然后是再通过L2归一化和点积，对Q和K进行余弦相似度计算。改进后的Adjust\_QKNorm具有零均值适应性，这能保证self-attention在计算矩阵相似性的时候除了考虑矩阵的方向，还能额外加入矩阵的数值大小信息。
%

We follow the idea of the QKNorm~\cite{henry2020query} and take into account that the insensitivity of the cosine similarity to values may lead to errors in the final results~\cite{sarwar2001item}, we propose Adjust\_QKNorm.
Notice that the matrices $Q$ and $K$  are three-dimensional vectors, the first dimension is the size of the batch, the second dimension is the length of the input sequence, and the third dimension is the representation vector of the embedding corresponding to the word. We subtract the values of the last dimension of both the matrices $Q$ and $K$  from their corresponding mean values (i.e., $\bar{Q}$ and $\bar{K}$), so that the sum of all elements on that dimension is equal to zero, followed by L2 normalization and dot product. Therefore, our proposed Adjust\_QKNorm has zero-mean adaptation, which ensures that self-attention can consider both the numerical difference information of the matrix and the direction of the matrix when calculating matrix similarity. The probability interval obtained by softmax after this calculation is reduced but does not change for the overall fluctuations, which may correct for some unreasonable scores.
The formula is shown as follows. 

\begin{equation}
\textbf{Q}=\frac{Q-\bar{Q}}{\|Q-\bar{Q}\|}, 
\textbf{K}=\frac{K-\bar{K}}{\|K-\bar{K}\|}
\end{equation}

\begin{equation}
\text { Attention }(Q, K, V)=\operatorname{softmax}\left(g * \textbf{Q} \textbf{K}^{T}\right) V
\label{formula:norm}
\end{equation}

Later, Transformer uses residual connection and layer normalization to make the operation dimension of  matrixes $X_{emb}$ and $X_{atten}$  consistent and normalize the hidden layer in the network to a standard normal distribution, which can accelerate the speed of the model training and convergence. In the fourth step, Transformer passes the feed-forward layer and two linear mapping layers. Then Transformer uses the activation function to generate the vector $X_{hidden}$. Later Transformer performs a residual connection and layer normalization to obtain the hidden semantic vector.

In the decoder part, Transformer uses the autoregressive mechanism, which predicts the next possible word in the generation based on the above content. Therefore, its next output needs to be inferred based on the output of the encoder part. Each decoder layer performs an attention operation on the final hidden state, which is the output of the encoder. The rest of the operation is identical to the encoder part. Finally, the output of the decoder $ht$ is sent to a fully connected neural network. This network can pass a $\operatorname{softmax}$ layer to predict the probability of the next token, which can be defined as follows.

\begin{equation}
P\left(y_{t+1} \mid y_{1}, y_{2}, \cdots, y_{t} \right)=\operatorname{softmax}\left(h_{t} W+b\right)
\end{equation}

We train our model parameters $\theta$ by minimizing the negative log likelihood of the target text $y$ for a given input text $x$. For the automatic shellcode generation task, $x$ means the natural language descriptions and $y$ means the code snippet. For the automatic shellcode summarization task, $x$ means the code snippet and $y$ means the natural language descriptions.
\begin{equation}
\mathcal{L}_{\theta}=-\sum_{i=1}^{|y|} \log P_{\theta}\left(y_{i} \mid y_{<i}, x\right)
\end{equation}

\subsection{Rule-based Repair Component}

For the automatic generation task of shellcode, we design a rule-based repair component to improve the performance of this task. 
An assembly statement in shellcode consists of the label, instruction, and operands. The format of the code segment is ``label: instruction  operand1, operand2  (options)". 
 After manual analysis of the cases with error output, 
We identified two types of errors: 
(1) global errors and (2) local errors. 
The uncontrollable nature of neural network construction may cause the low correlation problem between the generated code and ground-truth code, which  can be defined as a global error. 
While the local error mainly occurs  at the position of operand2. After analysis, we find that the local error is mainly caused by the OOV problem. Specially, the value of operand2 appears in the given comment but not in the vocabulary of the training set. 
Therefore, we mainly focus on fixing the local error type in this component. 
Hu et al.~\cite{huautomating} added the pointer network~\cite{see2017get} to the Transformer to solve this problem.
Yang et al.~\cite{yang2021comformer} alleviated the OOV problem with the help of sub-word tokenize method (i.e., Byte BPE~\cite{wang2020neural}). However, due to the nature of shellcode statements, it is more reliable  to  fix this type of error based on rules.
Specifically,
the values that appear in the operand position are divided into three categories (i.e., registers, addresses, and values). 
For example, if the input comment is ``ShellCodeGen: subtract 0x6374612e from the contents in ecx and save the result in ecx", the original output of the model is ``sub ecx, 0x1525152a", and we found there exits an inconsistency in the address. Therefore we extracted the correct address from the comment and used this address to update the generated shellcode. Then the fixed shellcode is  ``sub ecx, 0x6374612e".

%% file: 4setup.tex
\section{Experimental Setup}
\label{sec:setup}
In this section, we introduce the experimental subject, performance measures, baselines, and implementation details.

% In our empirical studies, we aim to investigate the following four research questions (RQs).

% \begin{itemize}
% \item \textbf{RQ1}: Can our proposed approach {\tool} outperform state-of-the-art baselines both in the shellcode generation task and the shellcode summarization task?
% \item \textbf{RQ2}: What is the effect of the rule-based repair component of {\tool} in the shellcode generation task?
% \item \textbf{RQ3}: What is the effect of the prefix and dual task learning of {\tool}  in shellcode generation and summarization tasks?
% \item \textbf{RQ4}: What is the effect of our proposed normalization method Adjust\_QKNorm both in shellcode generation and summarization tasks?

% \end{itemize}

\subsection{Experimental Subject}

In our empirical study, we choose the shellcode parallel corpus ShellCode\_IA32  shared by Liguori et al.~\cite{liguori2021shellcode_ia32} as our experimental subject. This corpus was gathered from shell-storm\footnote{\url{http://shell-storm.org/}} and Exploit Database\footnote{\url{https://www.exploit-database.net/}} in the period between 2000 and 2020. 
To ensure a fair comparison with the baselines, we followed
the data split method used by Liguori et al.~\cite{liguori2021shellcode_ia32}. 
Specifically, we divide the corpus into the training set (80\%), the validation set (10\%), and the test set (10\%). Then we can get the training set with 2,560 items, the validation set with 320 items, and the test set with 320 items respectively. Since our approach is to use only one model to learn the primal task and the dual task, we directly transform the corpus by exchanging the input and the output and distinguish the two tasks by adding the prefix. Thus we finally can get the training set with 5,120 items, the validation set with 640 items, and the test set with 640 items.
Table~\ref{tab:statistics} shows the statistical information of our used corpus.

\begin{table}[htbp] 
 \caption{Statistics of corpus used in our empirical study} 
 \label{tab:statistics}
 \vspace{-1mm}
 \begin{center}
 \begin{tabular}{cccccc} 
  \toprule 
  \multicolumn{5}{c}{
  \textbf{Statistics for Code Length}} \\ 
    \midrule 
Avg & Mode & Median & $<5$ & $<10$ & $<15$ \\ 
3.40 & 3 & 3 & $83.32\%$ & $97.30\%$ & $100\%$ \\ 
     \midrule 
  \bottomrule 
  \multicolumn{5}{c}{
  \textbf{Statistics for Comment Length}} \\ 
    \midrule 
Avg & Mode & Median & $<10$ & $<30$ & $<50$ \\ 
9.06 & 4 & 7 & $69.41\%$ & $98.13\%$ & $100\%$ \\ 
  \bottomrule 
 \end{tabular}
 \end{center}
 \vspace{-1mm}
\end{table}

\subsection{Performance Measures}

To quantitatively compare the performance between our proposed approach and the baselines, we choose the following four performance measures (i.e., BLEU, ROGUE, METEOR, and ACC), which have been widely used in previous neural machine translation and code generation studies~\cite{yang2020survey}. 
Notice the higher the value of the performance measure, the better the performance of the corresponding approach.

\noindent\textbf{BLEU.} BLEU (Bilingual Evaluation Understudy)~\cite{papineni2002bleu} is a variant of the precision measure. This performance measure can calculate the similarity by computing the $n$-gram precision of a candidate sentence to the
reference sentence, with a penalty for the  sentences with short length.

\noindent\textbf{ROUGE.}
ROUGE (Recall-Oriented Understudy for Gisting Evaluation)~\cite{lin2004rouge} is a set of measures for evaluating a candidate sentence and the reference sentence. In this study, we use ROUGE-L, where $L$ means LCS (Longest Common Subsequence). 
 
\noindent\textbf{METEOR.} METEOR (Metric for Evaluation of sentences with Explicit Ordering)~\cite{banerjee2005meteor} uses knowledge sources (such as WordNet) to expand the synset while taking into account the shape of words.

\noindent\textbf{ACC.} ACC (Exact Match Accuracy)~\cite{yin2017syntactic} is the fraction of exactly matching samples between the predicted output and the reference. Different from similarity-based measures, ACC considers whether the generated code is identical to the ground truth code. Therefore this measure can accurately evaluate the effectiveness of the generated shellcode for exploitability.

To ensure the implementation correctness of these performance measures, we utilize the nlg-eval library\footnote{\url{https://github.com/Maluuba/nlg-eval}}, which implements these performance measures for  natural language generation tasks.

\subsection{Baselines}

To show the competitiveness of our proposed approach {\tool},
we select six state-of-the-art baselines. Specifically, we classify these baselines into two groups. One group includes three approaches based on information retrieval (such as BM25~\cite{robertson1994some}, Jaccard~\cite{jaccard1912distribution}, and Levenshtein~\cite{levenshtein1966binary}). By observing our used corpus, we found that the baselines based on information retrieval are effective due to the consistency of comment styles and the presence of code cloning on this corpus.
% we considered these three  baselines since code cloning exists in shellcode development, which can be verified in our considered corpus. 
Another group includes three baselines based on deep learning, including NMT~\cite{liguori2021shellcode_ia32}, Dual Model~\cite{wei2019code}, and Transformer~\cite{vaswani2017attention}. 
Here, NMT is a recent state-of-the-art approach for the shellcode generation task~\cite{liguori2021shellcode_ia32}.
 Dual Model and Transformer are selected as our baselines since they are similar to our proposed approach. Moreover, these two baselines have shown competitive performance on both code generation and code summarization tasks~\cite{wei2019code,vaswani2017attention}. 

\subsection{Implementation Details}

In our study, we use Pytorch 1.6.0 to implement our proposed approach.
The hyper-parameters of our proposed approach  can be classified into three categories (i.e., the  hyper-parameters for the model structure, the hyper-parameters in the model training phase, and the hyper-parameters in the model test phase). These hyper-parameters and their value are shown in Table~\ref{Hyper-parameters}. Notice n\_layers means the number of layers (i.e., the depth of the model), and d\_model means the model size (i.e., the width of the model).

We run all the experiments on a computer with an Intel(R) Core(TM) i7-9750H 4210 CPU and a GeForce GTX1660ti GPU with 6 GB memory. The running OS platform is Windows 10.

\begin{table}[htbp]
 \caption{Hyper-parameters and their values in our empirical study}
 \vspace{-1mm}
 \begin{center}
\begin{tabular}{ccc}
\toprule
        \textbf{Category}               & \textbf{Hyper-parameter} & \textbf{Parameter Value} \\ \midrule
\multirow{4}{*}{Model Structure} & n\_layers       & 2     \\
                       & n\_heads        & 8     \\
                       & d\_model        & 256   \\
                       & hidden\_size    & 512   \\\midrule
\multirow{4}{*}{Model Training Phase} & dropout         & 0.25  \\
                       & optimizer       & Adam  \\
                       & learning rate   & 0.001 \\ 
                       & batch size      & 32    \\ \midrule
\multirow{1}{*}{Model Test Phase}  & beam size       & 3    \\
  \bottomrule
\end{tabular}
 \end{center}
 \vspace{-1mm}
 \label{Hyper-parameters}
\end{table}

%% file: 5result.tex
\section{Result Analysis}
\label{sec:results}

In this section, we perform result analysis for our designed four research questions.

\subsection{Result Analysis for RQ1}

\noindent\textbf{RQ1: Can our proposed approach {\tool} outperform state-of-the-art baselines both in the shellcode generation
task and the shellcode summarization task?}

In this RQ, we want to investigate how effective {\tool} is and how much performance improvement {\tool} can achieve over the state-of-the-art baselines. 
Since in the automatic shellcode generation task, we use the rule-based repair component. To guarantee a fair comparison between {\tool} and baselines, we also apply this component to our considered baselines.
% Specifically, in the automatic shellcode generation task, since we propose the Rule-based Repair Component, in order to fairly compare the performance between {\tool} and baselines, we add the Rule-based Repair Component also after the ShellCode generated by baselines, to repair the ShellCode generated by baselines.

We show the comparison results between our proposed approach {\tool} and baselines in Table~\ref{tab:RQ1}. 
For the shellcode summarization task,  
in terms of BLEU, {\tool} can improve the performance by 3.913\% to 12.380\%. 
In terms of ROUGE-L, {\tool} can improve the performance by 1.998\% to 13.438\%.
In terms of METEOR, {\tool} can improve the performance by 2.188\% to 6.102\%. 
For the shellcode generation task, 
in terms of BLEU, {\tool} can improve the performance by 8.013\% to 28.240\%. 
In terms of ROUGE-L, {\tool} can improve the performance by 3.132 \% to 15.472\%.
In terms of ACC, {\tool} can improve the performance by 9.0625 \% to 35.625\%.
Therefore, compared to the baselines, {\tool} can achieve significant performance improvement for both two tasks.

\begin{table}[htbp]
 \caption{The comparison results between our proposed approach {\tool} and baselines for two different tasks} 
 \label{tab:RQ1}
 \begin{center}
 \begin{tabular}{cccc} 
  \toprule 
  \multicolumn{4}{c}{
  \textbf{Automatic Shellcode Summarization}} \\ 
  \midrule 
    \textbf{Approach} & \textbf{BLEU-4 (\%)} & \textbf{ROUGE-L (\%)} & \textbf{METEOR (\%)} \\
    \midrule 
     BM25 & 45.097 & 53.325 & 32.604 \\ 
     Jaccard & 49.950 & 57.485 & 34.352 \\ 
     Levenshtein & 51.954 & 61.391 & 36.410 \\ 
     NMT & 47.662 & 62.408 & 34.756 \\ 
     Transformer & 51.474 & 64.765 & 35.891 \\ 
     Dual Model & 53.564 & 63.668 & 36.518 \\ 
     \textbf{{\tool}} & \textbf{57.477} & \textbf{66.763} & \textbf{38.706} \\ 
     \midrule 
  \bottomrule 
    \multicolumn{4}{c}{
  \textbf{Automatic Shellcode Generation}} \\ 
    \midrule 
     \textbf{Approach} & \textbf{BLEU-4 (\%)} & \textbf{ROUGE-L (\%)} & \textbf{ACC (\%)} \\
     \midrule 
     BM25 & 56.331 & 64.104 & 33.438 \\ 
     Jaccard & 60.429 & 66.872 & 34.688 \\ 
     Levenshtein & 49.374 & 64.260 & 27.813 \\ 
     NMT & 64.897 & 72.910 & 45.313 \\ 
     Transformer & 69.601 & 76.444 & 54.375 \\ 
     Dual Model & 68.718 & 75.423 & 51.563 \\ 
     \textbf{{\tool}} & \textbf{77.614} & \textbf{79.576} & \textbf{63.438} \\ 
  \bottomrule 
 \end{tabular}
 \end{center}
\end{table}

\begin{tcolorbox}[width=1.0\linewidth, title={}]
\textbf{Summary for RQ1:}
Our proposed approach {\tool} can outperform baselines in terms of four performance metrics in both the shellcode generation task and the shellcode summarization task.
\end{tcolorbox}
\vspace{-1mm}

\subsection{Result Analysis for RQ2}

\noindent\textbf{RQ2: What is the effect of the rule-based repair component of {\tool}
in the shellcode generation task?}

In this RQ, we want to investigate the effect of the rule-based repair component in the shellcode generation task. We compared the performance  of {\tool} and baselines before and after the usage of the rule-based repair component. As can be found in Table~\ref{tab:RQ2}, the performance of all baselines and {\tool} can be improved after using the rule-based repair component. For example, in terms of ACC, using this component can improve the performance by 8.44\% to 23.44\%.

\begin{table}[htbp]
 \caption{The results of {\tool} with/without the repair component} 
 \label{tab:RQ2}
 \begin{center}
 \begin{tabular}{cccc} 
  \toprule 
  \multicolumn{4}{c}{
  \textbf{Automatic Shellcode Generation Without The Repair Component}} \\ 
    \midrule 
     \textbf{Approach} & \textbf{BLEU-4 (\%)} & \textbf{ROUGE-L (\%)} & \textbf{ACC (\%)} \\
     \midrule 
     BM25 & 50.844 & 59.998 & 23.750 \\ 
     Jaccard & 51.343 & 59.930 & 17.188 \\ 
     Levenshtein & 37.396 & 54.101 & 4.375 \\ 
     NMT & 61.689 & 69.130 & 36.875 \\ 
     Transformer & 65.091 & 72.910 & 45.000 \\ 
     Dual Model & 62.992 & 71.119 & 40.000 \\ 
     \textbf{{\tool}} & \textbf{73.713} & \textbf{75.825} & \textbf{54.063} \\ 
     \midrule 
  \bottomrule 
    \multicolumn{4}{c}{
  \textbf{Automatic Shellcode Generation With The Repair Component}} \\ 
    \midrule 
     \textbf{Approach} & \textbf{BLEU-4 (\%)} & \textbf{ROUGE-L (\%)} & \textbf{ACC (\%)} \\
     \midrule 
     BM25 & 56.331 & 64.104 & 33.438 \\ 
     Jaccard & 60.429 & 66.872 & 34.688 \\ 
     Levenshtein & 49.374 & 64.260 & 27.813 \\ 
     NMT & 64.897 & 72.910 & 45.313 \\ 
     Transformer & 69.601 & 76.444 & 54.375 \\ 
     Dual Model & 68.718 & 75.423 & 51.563 \\ 
     \textbf{{\tool}} & \textbf{77.614} & \textbf{79.576} & \textbf{63.438} \\ 
  \bottomrule 
 \end{tabular}
 \end{center}
\end{table}

To show the competitiveness of our proposed repair component, we also consider other four methods for {\tool} to solve the OOV problem.
The first three methods (i.e., Byte BPE~\cite{wang2020neural}, Char BPE~\cite{sennrich2015neural}, and WordPiece~\cite{schuster2012japanese}) are based on sub-word tokenizatation. 
The last method  (i.e., pointer network~\cite{see2017get}) is based on the copy mechanism. The comparison results can be found in Table~\ref{tab:RQ2.5}. In this table, we can find the effectiveness of our designed simple repair component. Since sub-word tokenization relies on  large scale corpus, the three methods (i.e., Byte BPE~\cite{wang2020neural}, Char BPE~\cite{sennrich2015neural}, and WordPiece~\cite{schuster2012japanese}) do not perform well on our low-resource tasks. On the other hand, the pointer network~\cite{see2017get}  is based on probability for word reuse and this method still does not perform well on our tasks.

\begin{table}[htbp]
 \caption{The comparison results between our proposed repair component and other four methods for Automatic Shellcode Generation} 
 \label{tab:RQ2.5}
 \begin{center}
 \begin{tabular}{cccc} 
  \toprule 
  \multicolumn{4}{c}{
    \textbf{Automatic Shellcode Generation With Different Repair Methods}} \\ 
    \midrule 
     \textbf{Method} & \textbf{BLEU-4 (\%)} & \textbf{ROUGE-L (\%)} & \textbf{ACC (\%)} \\
     \midrule 
     Byte BPE & 65.682 & 72.114 & 49.063 \\ 
     Char BPE & 62.529 & 70.845 & 45.688 \\  
     WordPiece & 64.725 & 72.014 & 48.552 \\ 
     Pointer Network & 75.429 & 76.872 & 58.688 \\ 
     \textbf{{\tool}} & \textbf{77.614} & \textbf{79.576} & \textbf{63.438} \\ 
  \bottomrule 
 \end{tabular}
 \end{center}
\end{table}

\begin{tcolorbox}[width=1.0\linewidth, title={}]
\textbf{Summary for RQ2:}
The rule-based repair component is effective for the shellcode generation task and can significantly improve the performance of our proposed approach and baselines.
\end{tcolorbox}
\vspace{-1mm}

\subsection{Result Analysis for RQ3}

\noindent\textbf{RQ3:  What is the effect of the prefix and dual task learning of {\tool} in shellcode generation and summarization tasks?}

In this RQ, we want to investigate the effect of the prefix and dual task
learning in both  the shellcode generation task and the shellcode
summarization task. 
In this RQ, we consider two comparative approaches.
The first approach does not consider the prefix and we call this control approach ``Without Prefix".
The second approach does not consider the dual task learning and we call this control approach ``Single Task".
The comparison results between our proposed approach {\tool} and these control approaches can be found in Table~\ref{tab:RQ3}. 

\begin{table}[htbp]
 \caption{The comparison results between our proposed approach and the control approaches for two different tasks} 
 \label{tab:RQ3}
 \begin{center}
 \begin{tabular}{cccc} 
  \toprule 
  \multicolumn{4}{c}{
  \textbf{Automatic ShellCode Summarization}} \\ 
  \midrule 
    \textbf{Approach} & \textbf{BLEU-4 (\%)} & \textbf{ROUGE-L (\%)} & \textbf{METEOR (\%)} \\
    \midrule 
     Without Prefix & 50.277 & 64.371 & 38.088 \\ 
    %  with Prefix & \textbf{57.477} & \textbf{66.763} & \textbf{38.706} \\ 
     Single Task & 52.213 & 64.957 & 36.992 \\ 
     \textbf{{\tool}} & \textbf{57.477} & \textbf{66.763} & \textbf{38.706} \\ 
     \midrule 
  \bottomrule 
    \multicolumn{4}{c}{
  \textbf{Automatic ShellCode Generation}} \\ 
    \midrule 
     \textbf{Approach} & \textbf{BLEU-4 (\%)} & \textbf{ROUGE-L (\%)} & \textbf{ACC (\%)} \\
     \midrule 
     Without Prefix & 73.200 & 77.041 & 58.750 \\ 
    %  with Prefix & \textbf{77.614} & \textbf{79.576} & \textbf{63.438} \\
     Single Task & 66.192 & 76.152 & 56.250 \\ 
     \textbf{{\tool}} & \textbf{77.614} & \textbf{79.576} & \textbf{63.438} \\ 
  \bottomrule 
 \end{tabular}
 \end{center}
\end{table}

For the effect of the prefix in the shellcode summarization task, 
{\tool} can improve the performance by 7.2\%, 2.392\%, and 0.618\% respectively in terms of BLEU, ROUGE-L, and METEOR. 
For the effect of the prefix in the shellcode generation task, 
{\tool} can improve the performance by 4.414\%, 2.535\%, and 4.6875\% respectively in terms of BLEU, ROUGE-L, and METEOR.

For the effect of the dual task
learning in the shellcode summarization task, 
{\tool} can improve the performance by 5.264\%, 1.806\%, and 1.714\% respectively in terms of BLEU, ROUGE-L, and METEOR. 
For the effect of the dual task
learning in the shellcode generation task, 
{\tool} can improve the performance by 11.422\%, 3.424\%, and 7.1875\% respectively in terms of BLEU, ROUGE-L, and METEOR.

\begin{tcolorbox}[width=1.0\linewidth, title={}]
\textbf{Summary for RQ3:}
Both dual task learning and prefixing can have a positive impact on automatic shellcode generation and summarization tasks for {\tool}. 
\end{tcolorbox}
\vspace{-1mm}

\subsection{Result Analysis for RQ4}

\noindent\textbf{RQ4: What is the effect of our proposed normalization method Adjust\_QKNorm both in shellcode generation and summarization tasks?}

In this RQ, we want to investigate the impact of four different normalization methods on automatic shellcode generation and  summarization tasks. In particular, we consider the traditional normalization method in Transformer (i.e., PostNorm~\cite{ba2016layer}), two state-of-the-art normalization methods (i.e., PreNorm~\cite{xiong2020layer} and QKNorm~\cite{henry2020query}), and our proposed normalization method Adjust\_QKNorm. The details of these normalization methods are illustrated as follows.

\begin{itemize}
    \item PostNorm: In the original Transformer, Layer Norm follows Residual  is called PostNorm~\cite{ba2016layer}. The method PostNorm is very sensitive to parameters and needs to be carefully tuned to achieve promising results. For example, the essential warm-up learning rate strategy is often used. However, this strategy is very time-consuming.
  
    \item PreNorm: To alleviate the problems in PostNorm, PreNorm~\cite{xiong2020layer}  tries to put Layer Norm in the process of Residual, so that when the model is updated in reverse, the bottom parameters can directly obtain the information of the top gradient, which can effectively alleviate the issues of gradient vanishing and gradient explosion.
  
    \item QKNorm: To adopt Transformer for low-resource tasks, QKNorm~\cite{henry2020query} first calculates $Q$ and $K$ in the formula by L2 normalization the self-attention, and then adjust the dot product result of $Q$ and $K$ by a learnable parameter $g$ instead of dividing by $\sqrt{d_{k}}$.
  
    \item Adjust\_QKNorm: We follow the idea of the QKNorm~\cite{henry2020query} and propose the Adjust\_QKNorm. The QKNorm method can be found as replacing the calculation of the dot product of two vectors $Q$ and $K$ with the calculation of the cosine similarity of two vectors $Q$ and $K$. However, the insensitivity of cosine similarity to values can lead to errors in the results, therefore we improve the QKNorm by going through the zero-averaging operation and its formula can be found in Equation~(\ref{formula:norm}).
\end{itemize}

The comparison results between our proposed normalization method Adjust\_QKNorm and other normalization methods can be found in Table~\ref{tab:RQ4}. 
For the shellcode summarization task,  
in terms of BLEU, Adjust\_QKNorm can improve the performance by 3.027\% to 4.896\%. 
In terms of ROUGE-L, Adjust\_QKNorm can improve the performance by 1.357\% to 2.07\%.
In terms of METEOR, Adjust\_QKNorm can improve the performance by 0.164\% to 2.043\%. 
For the shellcode generation task with the repair component, 
in terms of BLEU, Adjust\_QKNorm can improve the performance by 0.123\% to 12.776\%. 
In terms of ROUGE-L, Adjust\_QKNorm can improve the performance by 0.521\% to 6.145\%.
In terms of ACC, Adjust\_QKNorm can improve the performance by 0.375\% to 12.1875\%.
Moreover, for the shellcode generation task without the repair component, Adjust\_QKNorm can also achieve the best performance.

\begin{table}[htbp]
 \caption{The comparison results between Adjust\_QKNorm and other normalization methods for two different tasks }
 \label{tab:RQ4}
 \begin{center}
 \begin{tabular}{cccc} 
  \toprule 
  \multicolumn{4}{c}{
  \textbf{Automatic Shellcode Summarization}} \\ 
  \midrule 
    \textbf{Method} & \textbf{BLEU-4 (\%)} & \textbf{ROUGE-L (\%)} & \textbf{METEOR (\%)} \\
    \midrule 
     PostNorm & 53.144 & 64.903 & 37.591 \\ 
     PreNorm & 52.581 & 64.693 & 36.663 \\ 
     QKNorm & 54.450 & 65.406 & 38.542 \\ 
     \textbf{Adjust\_QKNorm} & \textbf{57.477} & \textbf{66.763} & \textbf{38.706} \\
     \midrule 
  \bottomrule 
    \multicolumn{4}{c}{
  \textbf{Automatic Shellcode Generation Without The Repair Component}} \\ 
    \midrule 
     \textbf{Method} & \textbf{BLEU-4 (\%)} & \textbf{ROUGE-L (\%)} & \textbf{ACC (\%)} \\
     \midrule 
     PostNorm & 65.091 & 72.910 & 45.000 \\ 
     PreNorm & 60.551 & 69.642 & 41.250 \\ 
     QKNorm & 70.084 & 74.274 & 50.938 \\ 
     \textbf{Adjust\_QKNorm} & \textbf{73.713} & \textbf{75.825} & \textbf{54.063} \\ 
      \midrule 
  \bottomrule 
    \multicolumn{4}{c}{
  \textbf{Automatic Shellcode Generation With The Repair Component}} \\ 
    \midrule 
     \textbf{Method} & \textbf{BLEU-4 (\%)} & \textbf{ROUGE-L (\%)} & \textbf{ACC (\%)} \\
     \midrule 
     PostNorm & 71.765 & 78.108 & 60.313 \\ 
     PreNorm & 64.838 & 73.431 & 51.250 \\ 
     QKNorm & 77.491 & 79.055 & 63.063 \\ 
     \textbf{Adjust\_QKNorm} & \textbf{77.614} & \textbf{79.576} & \textbf{63.438} \\ 
  \bottomrule 
 \end{tabular}
 \end{center}
\end{table}

\begin{figure}[htbp]
	\centering
  \vspace{-1mm}
	\includegraphics[width=0.5\textwidth]{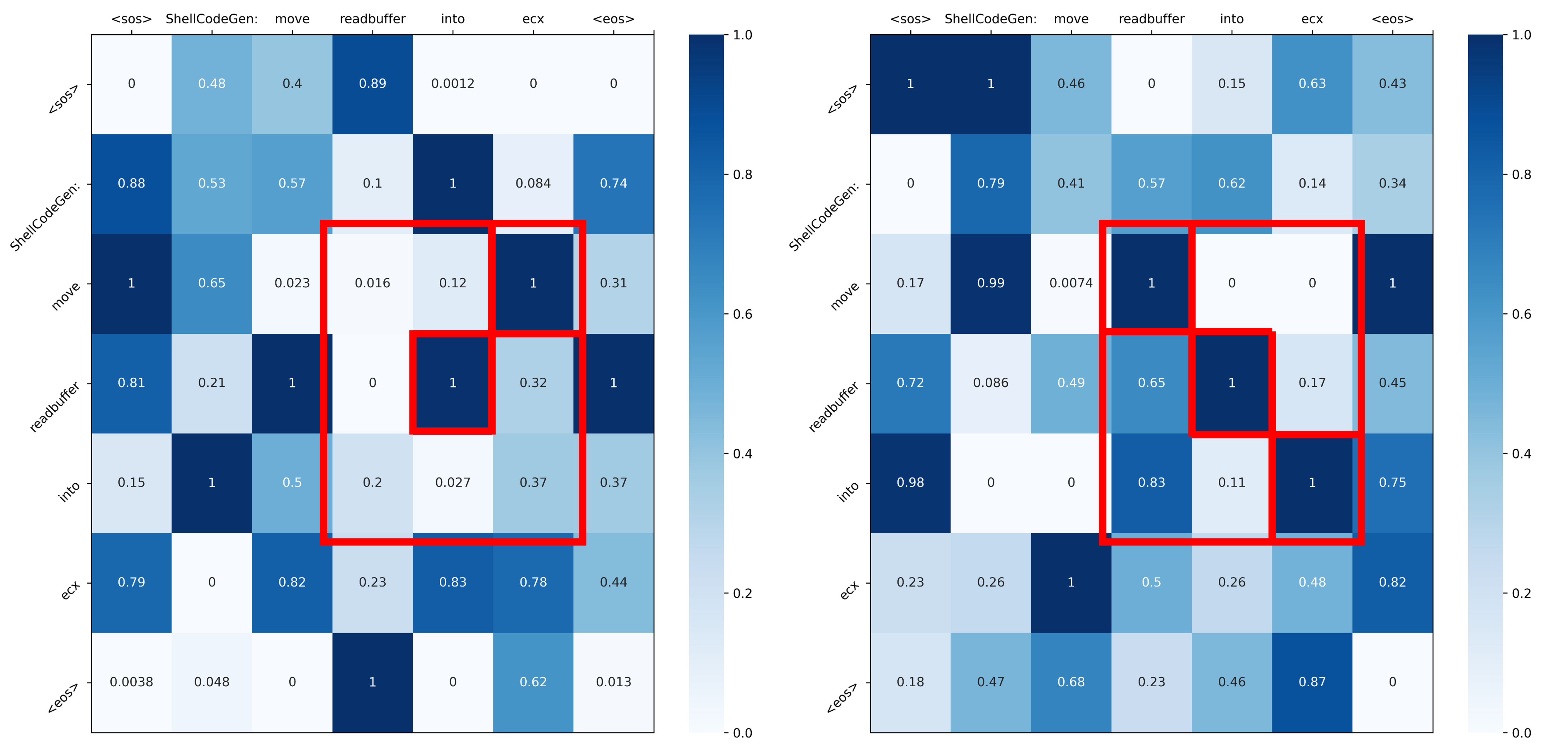}
	\caption{ Visualisation of the weight of self-attention by different normalization methods, the left subfigure is obtained by PostNorm, and the right subfigure is obtained by Adjust\_QKNorm}
  \vspace{-1mm}
	\label{fig:attention}
\end{figure}

For a more intuitive illustration of how the method Adjust\_QKNorm works, we selected an example for visualizing the weight of self-attention. As can be found in Fig.~\ref{fig:attention}, the input to the model is ``ShellCodeGen: move readbuffer into ecx", the left subplot represents the self-attentive visualization graph learned by the method PostNorm, and the right subplot  represents the self-attentive visualization graph learned by our proposed method Adjust\_QKNorm. In the left subplot, the method  PostNorm misaligned the attention for the input ``move readbuffer into ecx". While using the method Adjust\_QKNorm, the obtained attention can be corrected (i.e., ``move" corresponds to ``readbuffer", ``readbuffer" corresponds to ``into", and ``into" corresponds to ``ecx").

\begin{tcolorbox}[width=1.0\linewidth, title={}]
\textbf{Summary for RQ4:}
Our proposed normalization method Adjust\_QKNorm can achieve the best performance  on automatic
shellcode generation and summarization tasks for {\tool}.
% is more efficient compared to the normalisation method in the original Transformer and correctly corrects some of the incorrect self-attention values.
\end{tcolorbox}
\vspace{-1mm}

%% file: 6discuss.tex
\section{Discussions}
\label{sec:discuss}

In this section, we first perform sensitivity analysis on the hyperparameters of {\tool}. Then we conduct a human study on the shellcode summarization task. 

\begin{figure*}[htbp]
	\centering
  \vspace{-1mm}
	\includegraphics[width=1.0\textwidth]{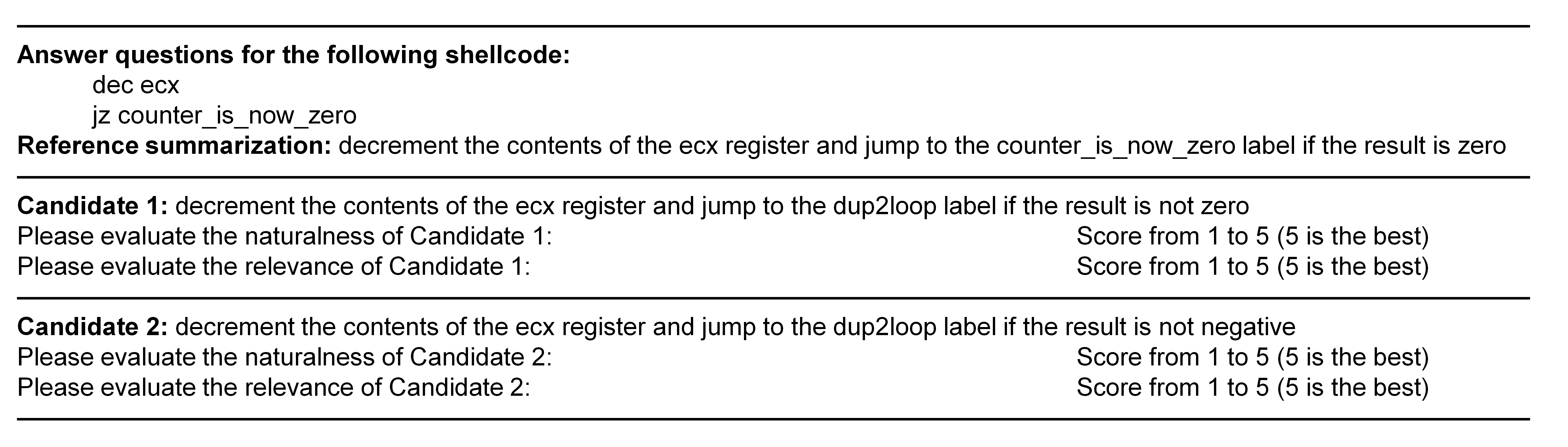}
	\caption{One page in the questionnaire of our human study}
  \vspace{-1mm}
	\label{fig:human}
\end{figure*}

\subsection{Sensitivity Analysis on the Hyperparameters}

In this subsection, we conduct experiments by performing sensitivity analysis on the parameters of {\tool}. We mainly focus on two parameters (i.e., the model size $d\_model$ and the number of layers $N$). The sensitivity analysis results on different parameters can be found in Table~\ref{tab:dis}. 
The optimal value of these parameters is set as follows: $d\_model$ is 256 and $N$ is 2.

\begin{table}[htbp]
  \begin{center}
 \caption{Sensitivity analysis on the hyperparameters (i.e., the model size and the number of layers)}
 \label{tab:dis}

%  \resizebox{0.5\textwidth}{!}{
\begin{tabular}{ccccc}
\toprule
\multicolumn{5}{c}{\textbf{Automatic Shellcode Summarization}}\\ \midrule
\multicolumn{1}{c|}{$N$}   & \multicolumn{1}{c|}{$d\_model$} & BLEU-4 (\%) & ROUGE-L (\%) & METEOR  (\%)\\ \midrule

\multicolumn{1}{c|}{
\multirow{4}{*}{1}} & \multicolumn{1}{c|}{128}  &   50.410 & 62.515 & 34.968 \\
\multicolumn{1}{c|}{} & \multicolumn{1}{c|}{256} &  52.827 & 65.070 & 37.768 \\
\multicolumn{1}{c|}{} & \multicolumn{1}{c|}{384} &  52.601 & 63.571 & 37.541 \\ 
\multicolumn{1}{c|}{} & \multicolumn{1}{c|}{512} &  55.785 & 66.025 & \textbf{38.940}\\\midrule
\multicolumn{1}{c|}{\multirow{4}{*}{2}} & \multicolumn{1}{c|}{128} & 55.821 & 64.944 & 37.694 \\
\multicolumn{1}{c|}{} & \multicolumn{1}{c|}{256} &\textbf{57.477}& \textbf{66.763} & 38.706 \\
\multicolumn{1}{c|}{} & \multicolumn{1}{c|}{384} & 54.837 & 64.596 & 37.993 \\ 
\multicolumn{1}{c|}{} & \multicolumn{1}{c|}{512} & 53.056 & 64.195 & 38.706 \\ \midrule
\multicolumn{1}{c|}{\multirow{4}{*}{3}} & \multicolumn{1}{c|}{128}   &     54.110 & 62.628 & 36.897 \\
\multicolumn{1}{c|}{}  & \multicolumn{1}{c|}{256}   & 46.813 & 60.976 & 33.816 \\
\multicolumn{1}{c|}{}  & \multicolumn{1}{c|}{384}   &    0  &  21.771    &   6.968 \\ 
\multicolumn{1}{c|}{}  & \multicolumn{1}{c|}{512}   &    0  &  21.416   & 6.763\\\midrule
\bottomrule 
\multicolumn{5}{c}{\textbf{Automatic Shellcode Generation}} \\ \midrule
\multicolumn{1}{c|}{$N$}   & \multicolumn{1}{c|}{$d\_model$} & BLEU-4 (\%) & ROUGE-L (\%) & ACC (\%)\\ \midrule

\multicolumn{1}{c|}{
\multirow{4}{*}{1}} & \multicolumn{1}{c|}{128}  &   64.362 & 71.748 & 48.438 \\
\multicolumn{1}{c|}{} & \multicolumn{1}{c|}{256} &  72.028 & 75.913 & 56.250 \\
\multicolumn{1}{c|}{} & \multicolumn{1}{c|}{384} &  73.235 & 78.214 & 59.688 \\ 
\multicolumn{1}{c|}{} & \multicolumn{1}{c|}{512} &  73.670 & 76.022 & 59.063\\ \midrule
\multicolumn{1}{c|}{\multirow{4}{*}{2}} & \multicolumn{1}{c|}{128} & 71.125 & 73.737 & 53.125 \\
\multicolumn{1}{c|}{} & \multicolumn{1}{c|}{256} &\textbf{77.614}& \textbf{79.576} & \textbf{63.438} \\
\multicolumn{1}{c|}{} & \multicolumn{1}{c|}{384} & 72.787 & 74.753 & 57.188 \\ 
\multicolumn{1}{c|}{} & \multicolumn{1}{c|}{512} & 70.373 & 75.757 & 55.313 \\ \midrule
\multicolumn{1}{c|}{\multirow{4}{*}{3}} & \multicolumn{1}{c|}{128}   &  74.520 & 73.360 & 53.438 \\
\multicolumn{1}{c|}{}  & \multicolumn{1}{c|}{256}   & 60.660 & 67.461 & 40.938 \\
\multicolumn{1}{c|}{}  & \multicolumn{1}{c|}{384}   &    0.303  &  15.539    &   1.250 \\ 
\multicolumn{1}{c|}{}  & \multicolumn{1}{c|}{512}   &    0.299  &  18.160  & 1.563\\  \bottomrule
\end{tabular}
% }
 \end{center}
\end{table}

The final results can be found in Table~\ref{tab:dis}. 
In this table, We find that it is harder to train the generation model when the value of $N$ (i.e., the depth of the model)   and the value of $d\_model$ (i.e., the width of the model) is large. 
Even if  {\tool} uses dual task learning and Adjust\_Norm to alleviate this problem,
the gradient vanishing issue tends to occur when training the generation model. 
Therefore we select small values for these parameters, which are more suitable for low-resource tasks in our study.

\subsection{Human Study on the Shellcode Summarization Task}

Different from the code generation task, the code summarization task requires human study to further verify the competitiveness of {\tool}. Stapleton et al.~\cite{stapleton2020human} showed that the performance measures (such as BLEU, ROUGE, and METEOR), which were commonly used for the code summarization task,  cannot necessarily capture how machine-generated code summaries actually affect program comprehension of developers. Therefore, we conduct the human study by following the methodology used in previous studies~\cite{gao2020generating,li2021secnn}. Since the Dual Model can achieve the best performance among all the baselines, we focus on analyzing the differences between comments generated by {\tool} and the baseline Dual Model. Due to the high cost of manually analyzing all these samples in the testing set, we use a commonly-used sampling method~\cite{singh2013elements} to select the minimum random samples. The number of the selected samples can be determined by the following formula:

\begin{equation}
MIN=\frac{n_{0}}{1+\frac{n_{0}-1}{size}}
\end{equation}
where $n_0$ depends on the selected confidence level and
the desired error margin $n_{0}\left(=\frac{Z^{2} \times 0.25}{e^{2}}\right)$. $Z$ is a confidence
level $z$ score and $e$ is the error margin. $size$ is the number of samples in the testing set. 
In our human study, we select $MIN$ examples with the
error margin $e$ = 0.05 at 95\% confidence level (i.e., $MIN =
175$).

We recruited six graduate students with rich development experience in shellcode to conduct our human study. We randomly selected 174 samples from the testing set, including ground-truth reference comment, the comments generated by {\tool} and Dual Model respectively. We generate a questionnaire for each master student and a page of a sample can be found in Fig.~\ref{fig:human}.
To guarantee a fair comparison, the order of the comments generated by these two approaches is random, which can make it difficult for students to know which approach these comments are generated by.
Here,  We mainly consider naturalness and relevance~\cite{gao2020generating} of the code summarization.
Specifically, naturalness refers to the grammatical correctness and fluency of the generated summarization. That is whether the generated code summarization is easy for developers to read and understand. 
Relevance refers to the relevance of the generated summarization to the target shellcode. That is whether developers can understand the intent of the shellcode through the code summarization.
The score ranges from 0 to 5 and 5 is the best value.

\begin{table}
  \begin{center}
 \caption{Results of our human study in terms of naturalness and relevance}
 \label{tab:human}
\begin{tabular}{ccc}
\toprule
\textbf{Type} & \textbf{{\tool}} & \textbf{Dual Model} \\
\midrule
Naturalness   & \textbf{4.12}            & 4.02                \\
Relevance     & \textbf{4.08}            & 3.95               \\
\bottomrule
\end{tabular}
 \end{center}
\end{table}

After our human study, we compute the average score of the six students' feedback and the results are shown in Table~\ref{tab:human}. 
% Finally, we calculated the average of the six volunteers' feedback and the results are shown in Table 11. 
In this table, We can find that {\tool} can outperform the baseline Dual Model by 0.10 and 0.13 respectively when considering  naturalness and relevance. Therefore, our human study can  further verify the competitiveness of {\tool}.

%% file: 7threats.tex
\section{Threats to validity}
\label{sec:threat}

In this section, we mainly analyze the potential threats to the validity of our empirical study.

\noindent\textbf{Internal Threats.} 
The first internal threat is the potential defects in the implementation of {\tool}. 
To alleviate this threat, we check our implementation carefully and use mature libraries (such as PyTorch and TorchText\footnote{\url{https://pytorch.org}}).
The second internal threat is the implementation of our considered baselines.
To alleviate this threat, we use textdistance library\footnote{\url{https://github.com/life4/textdistance}} to implement Jaccard and Levenshtein, we use rank\_bm25\footnote{\url{https://github.com/dorianbrown/rank_bm25}} library to implement BM25, we use OpenNMT-py~\cite{klein-etal-2017-opennmt} library\footnote{\url{https://github.com/OpenNMT/OpenNMT-py}} to implement NMT and Transformer, and we implement Dual Model based on the code shared by Wei et al\footnote{\url{https://github.com/code-gen/cscg}}.

\noindent\textbf{External Threats.} 
The main external threat is the choice of the corpus. To alleviate this threat, we select the popular corpus provided by Liguori et al.~\cite{liguori2021shellcode_ia32} and this corpus was gathered from shell-storm2 and Exploit Database. 
% In the future, we want to gather more corpora from other commercial or open-source projects and verify the effectiveness of our proposed method {\tool}.

\noindent\textbf{Construct Threats.} 
The construct threat in this study is the performance measures used to evaluate the performance of {\tool}.
To alleviate these threats, we first choose the similarity-based performance measures (such as BLEU and Rouge-L) for the shellcode generation task, which have been used by the previous code generation studies. However, these measures are not designed by running the exploits. Therefore the generated shellcode might not necessarily offer a remote shell. Therefore we further consider the ACC (Exact Match Accuracy) to verify the correctness of the generated shellcode. For the shellcode summarization task, we choose three performance measures (i.e., BLEU, Rouge-L, and METEOR), which have been used by the previous code summarization studies~\cite{hu2018deep}\cite{yang2021comformer}\cite{ahmad2020transformer}. 

% \noindent\textbf{Conclusion Threats.} 
% The conclusion threat in our study is we do not perform cross-validation (CV) in our research.
% In our study, the data split on the corpus is consistent with the experimental setting in the previous pseudo-code generation study~\cite{oda2015learning}.
% This can guarantee a fair comparison with the baselines (i.e., the model construction and prediction are on the same training set, the validation set, and the testing set).
% Using CV can comprehensively evaluate our proposed method {\tool}, since different split may result in a diverse training set, validation set, and testing set. However, this model evaluation method has not been commonly used for neural machine translation due to high training computational cost. 
% Besides, The proposed code feature extractor in this study has only been improved for the pseudo-code generation task and has not been attempted to be validated on other short text translation applications.

%% file: 8conclusion.tex
\section{Conclusion}
\label{sec:conclusion}

In this study, we propose a novel approach {\tool} via Transformer and dual learning to solve the automatic shellcode generation
and summarization tasks. Specifically, we formalize automatic shellcode generation and summarization as dual tasks, using a shallow Transformer for learning, while proposing Adjust\_QKNorm to adapt our proposed approach on low-resource tasks. Finally, to mitigate the OOV problem, we propose a rule-based repair component to improve the accuracy of {\tool} on automatic shellcode generation.
Empirical results verified the effectiveness of our proposed approach.
Then, we analyze the 
effectiveness of the component settings
in {\tool}. Finally, we also conduct a human study to verify
the effectiveness of {\tool}.
% In the future, we want to improve the performance of our proposed approach {\tool} by considering more advanced normalization methods and pre-training models. 
% Moreover, we also want to verify the effectiveness of {\tool} by considering corpora gathered from other commercial or open-source projects developed by other programming languages. 